%% file: paper.tex
\documentclass[11pt]{article}

\usepackage[top=1.0in,bottom=1.0in,left=1.0in,right=1.0in]{geometry}

\usepackage{url}
\usepackage{listings}
\usepackage[font=footnotesize,labelfont=bf]{caption}
\usepackage{wrapfig}
\lstdefinelanguage{Scala}{
  morekeywords={
          abstract,case,catch,class,def,do,else,extends,
          false,final,finally,for,forSome,if,implicit,import,lazy,
          match,new,null,object,override,package,private,protected,
          return,sealed,super,this,throw,trait,true,try,type,
          val,var,while,with,yield},
  otherkeywords={=>,<-,<\%,<:,>:,\#,@},
  sensitive=true,
  morecomment=[l]{//},
  morecomment=[n]{/*}{*/},
  morestring=[b]",
  morestring=[b]',
  morestring=[b]"""
}[keywords,comments,strings]
\newcommand{\var}[1]{\text{\lstinline+#1+}}

\input{head}

\renewcommand{\paragraph}[1]{\vspace{.6em}\noindent\textbf{#1}\hspace{.5em}}

\newcommand\toolName{Scala{\sc Proust}}

\title{Proust: A Design Space for \\
  Highly-Concurrent Transactional Data Structures}
\date{}
\author{
  Thomas Dickerson\\
  Computer Science Department\\
  Brown University\\
  \texttt{thomas\_dickerson@brown.edu}
  \and
  Paul Gazzillo\\
  Computer Science Department\\
  Yale University\\
  \texttt{paul.gazzillo@yale.edu}
  \and
  Maurice Herlihy\\
  Computer Science Department\\
  Brown University\\
  \texttt{mph@cs.brown.edu}
  \and
  Eric Koskinen\\
  Computer Science Department\\
  Yale University\\
  \texttt{eric.koskinen@yale.edu}
}
\begin{document}

\maketitle 
\begin{abstract}
Most STM systems are poorly equipped to support libraries of concurrent data structures.
One reason is that they typically detect conflicts by tracking transactions' \emph{read sets} and \emph{write sets},
an approach that often leads to false conflicts.
A second is that existing data structures and libraries often need to
be rewritten from scratch to support transactional conflict detection
and rollback.
This paper introduces \emph{Proust},
a framework for the design and implementation of transactional data structures.
Proust is designed to maximize re-use of existing well-engineered
by providing transactional ``wrappers'' to make existing
thread-safe concurrent data structures transactional.
Proustian objects are also integrated with an underling STM system,
allowing them to take advantage of well-engineered STM conflict detection
mechanisms.
Proust generalizes and unifies prior approaches such as boosting and predication.
\end{abstract}

\input{intro}
\input{overview}
\input{conflictabstraction}

\input{shadow}
\input{opacity}
\input{implementation}
\input{experiment}

\input{relwork}

\section{Conclusions}
We believe that the Proust methodology serves as useful niche in the transactional data structures ecosystem.
Like Boosting, we offer sufficient expressivity to wrap arbitrary data structures, but with reduced design complexity (constraints are expressed as commutativity of updates to abstract state elements, rather than as pairwise commutativity rules between operations).
Furthermore, our well-characterized design-space permits the use of different synchronization and update strategies to selectively optimize the performance of wrapped data structures for different STMs and different expected work-loads.

Benchmarking shows that we outperform, or are competitive with, pure STM solutions, and leveraging existing data structures libraries allows us to achieve that performance without the cost of writing new implementations from scratch. While we are outperformed by Predication on the map throughput tests, we believe that our utility as a tool for wrapping arbitrary data structures will encourage use beyond sets and maps.

We see several avenues for future work. 
On the theoretical front, an extension of our log-combining optimization from memoized replays to snapshot replays and undo logs would further improve performance.
Further development of the conflict abstraction tools based on SAT/SMT could allow wrappers to be generated automatically and allow conservative estimates of commutativity to be parameterized by the size of a condensed state-space.
Additionally, while we have presented two possible implementations of shadow copy functionality, it's possible that some concurrent data-structures might be designed from the ground up to calculate the results of speculative operations to provide similar functionality natively.
On the implementation front, the Proust methodology could be implemented as a framework for other STMs.

\vfill
\pagebreak
\bibliographystyle{abbrv}
\bibliography{tm,biblio,ejk}

\pagebreak
\appendix

\pagebreak
\input{proofs}

\pagebreak
\section{Code Listings - Abstract Lock}
\begin{lstlisting}[language=Scala,caption=The public API for AbstractLock.,label=lst:abstractLockAPI]
class AbstractLock[Key](lAP:LockAllocatorPolicy[Key], updStrat:UpdateStrategy) {
  sealed trait LockFor{ def key:Key; def write:Boolean }
  case class Write(override val key:Key) extends LockFor{override val write = true}
  implicit class Read(override val key:Key) extends LockFor{override val write = false}
  
  def apply[Z](acquire:LockFor*)(f: =>Z)(invF:Z=>Unit = null)(implicit txn:InTxn):Z
}
\end{lstlisting}
\section{Code Listings - Concurrent Map Trait}
\begin{lstlisting}[language=Scala,caption=A transactional Map trait. Size has been reified out of the abstract state as an optimization.,label=lst:txnMapTrait]
trait MapTrait[K, V] {
  protected def lAP:LockAllocatorPolicy[K]
  protected def uStrat:UpdateStrategy
  protected val abstractLock = new AbstractLock(lAP, uStrat)
  protected var committedSize = Ref(0).single
  
  def put(k : K, v : V)(implicit txn:InTxn) : Option[V]
  def get(k : K)(implicit txn:InTxn) : Option[V]
  def contains(k : K)(implicit txn:InTxn) : Boolean
  def remove(k : K)(implicit txn:InTxn) : Option[V]
  def size(implicit txn:InTxn):Int = committedSize()
}
\end{lstlisting}
\section{Code Listings - Concurrent Priority Queue Trait}
\begin{lstlisting}[language=Scala,caption=A transactional PQueue trait with two abstract state elements.,label=lst:txnPQueueTrait]
object PQueueTrait {
	sealed trait PQueueState
	case object PQueueMin extends PQueueState
	case object PQueueMultiSet extends PQueueState
}

trait PQueueTrait[V] {
	protected def lAP:LockAllocatorPolicy[PQueueTrait.PQueueState]
	protected def uStrat:UpdateStrategy
	protected val abstractLock = new AbstractLock(lAP, uStrat)
	
	def insert(v : V)(implicit txn:InTxn) : Unit
	def min(implicit txn:InTxn):Option[V]
	def contains(v : V)(implicit txn:InTxn) : Boolean
	def removeMin(implicit txn:InTxn) : Option[V]
	def size(implicit txn:InTxn):Int
}
\end{lstlisting}

\pagebreak
\input{smt}

\end{document}

%% file: head.tex
\usepackage{changepage}
\usepackage{titlesec}
\usepackage{subcaption}
\usepackage{amsmath}
\usepackage{amssymb}
\usepackage{wrapfig}
\usepackage{tikz}
\usepackage{mathabx}
\usepackage{qtree}
\usepackage{harpoon}
\usepackage{graphicx}
\usepackage{epstopdf}
\usepackage{epsfig}
\usepackage{color}
\usepackage{program}
\usepackage{multirow}
\usepackage{proof}
\usepackage{url}
\usepackage{hyperref}
\usepackage{stmaryrd}
\usepackage{fancyvrb}

\newcommand\Loadedframemethod{default}
\usepackage[framemethod=\Loadedframemethod]{mdframed}

\titlespacing*{\section}
{0pt}{0.3cm}{0.2cm}
\titlespacing*{\subsection}
{0pt}{0.3cm}{0.2cm}

\titlespacing*{\paragraph}
{0pt}{0.2cm}{0.1cm}

\usepackage{listings}
\usepackage{lstlinebgrd}
\let\origthelstnumber\thelstnumber
\makeatletter
\newcommand*\Suppressnumber{%
  \lst@AddToHook{OnNewLine}{%
    \let\thelstnumber\relax%
     \advance\c@lstnumber-\@ne\relax%
    }%
}

\newcommand*\Reactivatenumber{%
  \lst@AddToHook{OnNewLine}{%
   \let\thelstnumber\origthelstnumber%
   \advance\c@lstnumber\@ne\relax}%
}

\lstdefinestyle{nonumbers}
               {numbers=none}

\input{defs}

\input{genericmacros}

\input{specificmacros}
\usepackage{amsthm}
\newtheorem{theorem}{Theorem}[section]
\newtheorem{definition}{Definition}[section]

\usepackage{algorithm}
\usepackage{algpseudocode}
\algrenewcommand{\algorithmiccomment}[1]{{\small\hfill$\triangleright$ #1}}
\algrenewcommand\algorithmicindent{1em}

\algnewcommand\algorithmicswitch{\textbf{switch}}
\algnewcommand\algorithmiccase{\textbf{case}}
\algnewcommand\algorithmicassert{\texttt{assert}}
\algnewcommand\Assert[1]{\State \algorithmicassert(#1)}%
\algdef{SE}[SWITCH]{Switch}{EndSwitch}[1]{\algorithmicswitch\ #1\ \algorithmicdo}{\algorithmicend\ \algorithmicswitch}%
\algdef{SE}[CASE]{Case}{EndCase}[1]{\algorithmiccase\ #1}{\algorithmicend\ \algorithmiccase}%
\algtext*{EndSwitch}%
\algtext*{EndCase}%
\algtext*{EndFor}%
\algtext*{EndIf}%

\newenvironment{itemize*}%
  {\begin{itemize}%
    \setlength{\itemsep}{0.0in}%
    \setlength{\topsep}{0.0in}%
    \setlength{\parskip}{0.0in}}%
  {\end{itemize}}
\newenvironment{enumerate*}%
  {\begin{enumerate}%
    \setlength{\itemsep}{0.0in}%
    \setlength{\topsep}{0.0in}%
    \setlength{\parskip}{0.0in}}%
  {\end{enumerate}}

\definecolor{light-gray}{gray}{0.85}

\newcommand\red[1]{{\color{red} #1}}

\newcommand\ignore[1]{}

\newcommand\eg{{\it e.g.}}

\newcommand\FULL[1]{\red{(omitted content)}}


%% file: defs.tex
\newcommand\APPLY{{\sc apply}}
\newcommand\UNAPP{{\sc unapply}}
\newcommand\PUSH{{\sc push}}
\newcommand\UNPUSH{{\sc unpush}}
\newcommand\PULL{{\sc pull}}
\newcommand\UNPULL{{\sc unpull}}
\newcommand\CMT{{\sc cmt}}



%% file: genericmacros.tex
\usepackage{ifthen}
\newcommand{\DONE}[1]{}
\newcommand{\COMMENT}[1]{}

\newcounter{programlinenumber}

\newboolean{TR}
\setboolean{TR}{false}
\ifthenelse{\boolean{TR}}{

\newcommand{\TrOnly}[1]{#1}
\newcommand{\SubOnly}[1]{}
\newcommand{\TrOnlyInFootnote}[1]{#1}
\newcommand{\TrOnlyInTable}[1]{#1}}
{

\newcommand{\TrOnly}[1]{}
\newcommand{\SubOnly}[1]{#1}
\newcommand{\TrOnlyInFootnote}[1]{}
\newcommand{\TrOnlyInTable}[1]{}}



%% file: specificmacros.tex
\newcommand{\hiddentext}[1]{}

\renewcommand{\phi}{\varphi}

\definecolor{graybck}{gray}{0.9}

%% file: intro.tex
\section{Introduction}
Software Transactional Memory (STM) has become a popular alternative
to conventional synchronization models,
both as programming language libraries \cite{HarrisMJH08,IntelCpp,Goodman2013150,deuce}
and as stand-alone systems~\cite{
ali-reza:2007:ppopp,
Dalessandro:2010:NSS:1693453.1693464,
gottschlich:invalidation:cgo:2010,
Herlihy:2003:podc,
marathe:2008:ppopp,
Riegel:2007:TTM:1248377.1248415}.
STM systems structure code as a sequence of \emph{transactions},
blocks that are executed \emph{atomically},
meaning that steps of concurrent transactions do not appear to interleave.

Most STM systems, however,
are not well-equipped to support libraries of concurrent data structures.
One limitations is that STM systems typically detect conflicts by tracking
transactions' \emph{read sets} and \emph{write sets},
an approach that often leads to false conflicts,
when operations that could have correctly executed concurrently are
deemed to conflict, causing unnecessary rollbacks and serialization.
Instead, it would be preferable to exploit data type semantics to
enhance concurrency by recognizing when operations do not interfere
at the semantic level, even if they might interfere at some lower level.

A second limitation is that existing thread-safe libraries and data structures
must typically be rewritten from scratch to accommodate idiosyncrasies
of the underlying STM system.
The prospect of discarding so much carefully engineered concurrent
software is unappealing.
Instead, it would be preferable to provide a pathway for porting
at least some existing thread-safe concurrent data structures and
algorithms into STM systems.

We are not the first to identify these limitations.
\emph{Transactional boosting}~\cite{HerlihyK2008} describes a
methodology for constructing a transactional  ``wrapper'' for prior
thread-safe concurrent data structures.
A boosting wrapper requires identifying which operations commute,
as well as providing operation inverses.
Boosting is a stand-alone process, not integrated with an STM.
\emph{Transactional predication}~\cite{BronsonCCOn2010} describes a
way to leverage standard STM functionality to encompass
highly-concurrent sets and maps.
Predication, however, does not appear to extend beyond sets and maps,
and does not provide an explicit path to migrate legacy data
structures and libraries.
\emph{Software transactional objects}~\cite{HermanIHTKLS2016} (STO)
is an STM design that provides built-in primitives to track conflicts among
arbitrary operations, not just read-write conflicts.
It does not provide a migration path for legacy libraries.

This paper introduces \emph{Proust}\footnote{This name is a portmanteau of \emph{predication} and \emph{boosting}, both influential prior works. The name is also an \emph{hommage} to Marcel Proust, an author famous for his exploration of the complexities of memory.},
a framework which generalizes key insights from transactional boosting and predication.
Proust is designed to ease re-use of existing well-engineered
libraries in two ways.
First,
Proust is a methodology for the design and implementation of
transactional ``wrappers'' that transform existing libraries of
thread-safe concurrent data structures into transactional data
structures so as to minimize false data conflicts.
(Unlike predication,
Proust supports objects of arbitrary abstract type, not just sets and maps.)
Second, unlike boosting,
Proustian objects are integrated with an underling STM system,
allowing them to take advantage of well-engineered STM conflict detection
mechanisms.

Two key elements are necessary to wrap an existing non-transactional
concurrent data structure into an STM-compatible object.
First, as with boosting, it is necessary to characterize the
commutativity relationships between the various operations on that
data structure.
For example, in a map, queries and updates to non-intersecting key
ranges commute. 
Sometimes, this determination can be performed automatically by
reduction to SMT solvers, as discussed below in
Section~\ref{section:conflict}. 
Moreover, in most cases these relationships can be conservatively
approximated with traditional two-phase locks, without much loss in
expressivity. 

Second, the wrapper must provide an \emph{update strategy}, 
either by providing an inverse for each operation,
or a \emph{shadow copy}\footnote{A shadow copy essentially provides copy-on-write
  semantics. The most effective way to provide this functionality is
  type-dependent, and we discuss detailed examples in
  Section~\ref{section:shadow}.} functionality.
Inverses and shadow copies correspond to alternative update strategies,
as discussed in Section~\ref{section:designspace}. 

This paper makes the following contributions:
\begin{itemize*}
\item Novel ways to transform black-box highly-concurrent linearizable
  data structures into transactional objects in a way that minimize false conflicts,
  generalizing key insights of boosting and predication.
\item The concept of a \emph{conflict abstraction} that translates
an abstract data type's semantic notions of conflict
into concrete forms that can be efficiently managed by a generic
software transactional memory run-time (Section~\ref{section:conflict}).
\item
Systematic guidelines for choosing a transactional API for an
underlying thread-safe data structure.
Effectively, the transactional API must choose between
\emph{optimistic} or \emph{pessimistic} conflict resolution, and
\emph{eager} or \emph{lazy} update strategies
(Section~\ref{section:designspace}) 
\item The \toolName{} prototype implementation, built on top of ScalaSTM,
  shows scalability matching existing specialized approaches,
  but with a much wider range of applicability.
(Section~\ref{section:exp})
\end{itemize*}

%% file: overview.tex
\section{Overview}
\label{section:designspace}
\paragraph{The Proust Design Space\hfill}
\begin{wrapfigure}[29]{r}{2.5in}
\begin{tabular}{c}
\includegraphics[width=2.5in]{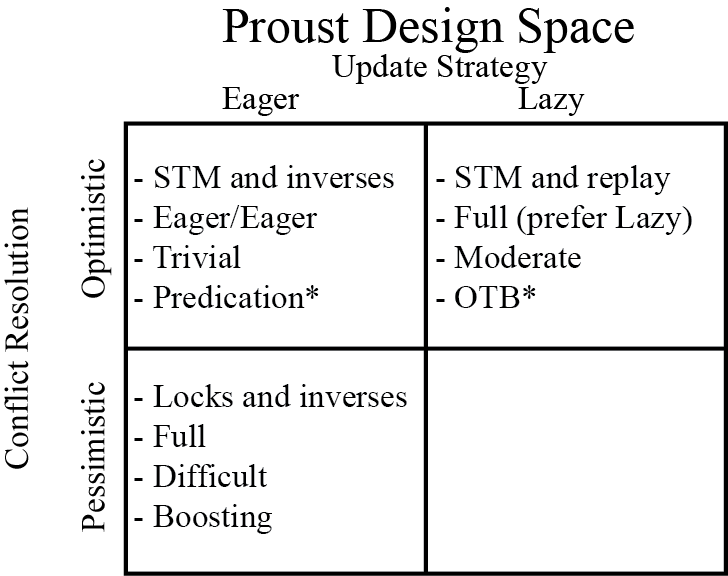} \\
\includegraphics[width=2.5in]{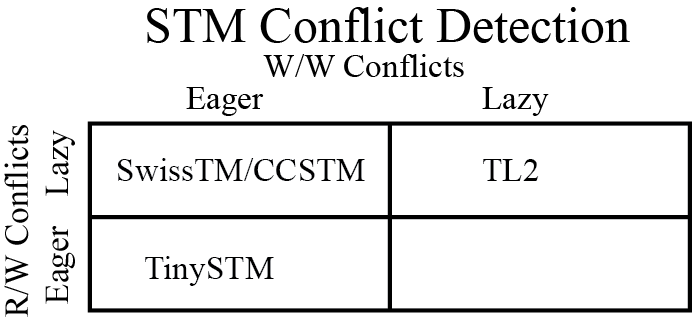}
\end{tabular}
\caption{\label{fig:designspaces} Design spaces for STMs and STM-integrated data structures.  The right-hand table maps conflict detection strategies to popular STMs as outlined by Dragojevic, et al~\cite{dragojevic2009stretching}. The left-hand table outlines the Proust design space, listing for each how the transactional API is implemented, its compatibility with the STM strategies at right, the difficulty of correctly synchronizing an AbstractLock implementation with the underlying STM, and the most conceptually similar prior work.}
\end{wrapfigure}
The Proust methodology,
like Boosting, Predication, and optimistic transactional boosting
(OTB)~\cite{HerlihyK2008, BronsonCCOn2010,  Hassan2014OTB} is based on
the principles that 
(1) synchronization conflicts should be defined over an object's abstract (not concrete) state,
and (2) the abstract state can be mapped to an underlying STM mechanisms by proper
synchronization and an ability to roll back changes.

By the Proust \emph{design space},
we mean the following several implementation choices.
Concurrency control can be optimistic (as in Predication, OTB)
or pessimistic (as in Boosting).
The base data structure can be modified \emph{eagerly} as the transaction executes,
or \emph{lazily} postponed to commit time.
Each prior work cited commits to one fixed choice from each category,
while Proust provides a unifying structure allowing choices to be mixed and matched.

\paragraph{The Proust Methodology}
Proust detects and resolves synchronization conflicts through
\emph{conflict abstractions},
which are (roughly speaking) maps carrying abstract states to
concurrency control primitives provided by an underlying STM.
At the concrete end,
programmers are responsible for providing a \emph{lock allocator policy} (LAP),
which allocates concurrency control primitives as needed.
The LAP is either optimistic or pessimistic.
A pessimistic LAP allocates standard re-entrant read-write locks,
while an optimistic LAP returns an object which maps \texttt{lock}
invocations to operations on standard STM memory locations,
allowing the STM to detect and manage synchronization conflicts.

Programmers also choose whether to wrapped objects are modified lazily or eagerly.
A lazy strategy requires the ability to construct a shadow copy
(Section~\ref{section:shadow}),
while an eager strategy requires each operation have a declared inverse,
registered as a rollback handler by the abstract lock.
If shadow copy functionality is provided,
each operation on the wrapped object is forwarded through a \emph{replay log}.
The replay log computes the result of the operation at execution time
using the shadow copy,
and registers a handler that to reapply the operation to the wrapped object.  

There are many considerations in making these choices,
depending on the data structure's operations,
or the strengths and weaknesses of the underlying STM system
(Figure \ref{fig:designspaces}).

Not all combinations make sense.
For example,
the empty quarter in Figure~\ref{fig:designspaces} reflects an
impractical combination of choices.
Some combinations are more complicated.
For example,
as discussed in Section~\ref{section:opacity},
the eager-optimistic combination satisfies \emph{opacity}~\cite{opacity} only
under STMs that provide eager detection of both read-write and
write-write conflicts. 
Here, Proust differs from Predication.
While both are eager,
Predication delegates state modifications to the underlying STM,
instead of using the STM only for synchronization.
Some second-order considerations include the degree to which the STM's
contention management is exposed and can be coupled with traditional
pessimistic locks,
and the memory overhead of allocating shadow copies on target systems. 

%% file: conflictabstraction.tex
\newcommand{\incr}{\var{incr()}}
\newcommand{\decr}{\var{decr()}}

\section{Conflict Abstraction}
\label{section:conflict}
The principal challenge for any type-specific transactional object
implementation is how to map type-specific notions of
conflict into a low-level synchronization framework,
a notion we call \emph{conflict abstraction}.
Like others~\cite{BronsonCCOn2010,HerlihyK2008,Koskinen:2010,KP:PLDI2015},
we identify type-specific synchronization conflicts with \emph{failure to commute}.
(Two operations commute if applying them in either order yields the
same return values and the same final object state.)

Transactional boosting focuses on making
thread-safe linearizable data structures transactional.
Boosting provides conflict abstraction by mapping commutativity-based
conflicts to \emph{abstract locks},
exclusive locks acquired explicitly by transactions before calling
base object operations and released implicitly on commit or abort.
Locks are assigned so that operations that do not commute, conflict.
For example,
a hashmap might associate an abstract lock with each key value (or its hash).
Hashmap operations \var{get}(5) and \var{put}(6,\var{foo}) commute,
and would acquire distinct, non-conflicting locks.
Operations on the same key do not commute,
and would contend for the same lock,
delaying one operation while the other's transaction is active.
Boosting is \emph{pessimistic} in the sense that synchronization
occurs before applying an operation,
but there are optimistic variations~\cite{Hassan:2014:OTB:2555243.2555283,Koskinen:2010,KP:PLDI2015}.

Transactional Predication~\cite{BronsonCCOn2010}
focuses on constructing highly-concurrent transactional sets and maps compatible with an underlying STM.
Instead of mapping conflicts to abstract locks,
it maps them to memory locations synchronized by the STM
using read-write synchronization.
This kind of conflict abstraction consists of
(1) a memory region \var{mem} whose synchronization and recovery is
managed by the underlying STM, 
(2) a non-transactional thread-safe map that links keys to unique
memory locations within that region.
For example,
to store a new key-value pair $k,v$ in the transactional map,
allocate an unused index $m$ into the STM-managed region,
non-transactionally bind $k$ to $m$ in the non-transactional map,
and transactionally store $v$ in \var{mem}$[m]$.
Conflicting map operations become conflicting read or write
operations on \var{mem}.
These conflicts are detected and resolved by the STM.

We now propose to unify and generalize these approaches.
As in Predication, we start  with an underlying STM,
and allocate an array of STM-managed memory locations \var{mem}
of size $M$, a parameter to be tuned later.
The STM provides synchronization and recovery for transactions that
read and write these locations.
A conflict abstraction assigns to each operation of
abstract type one or more memory locations to be read or written
is such a way that non-commuting operations trigger conflicting memory accesses.
It is important to note that these STM read-write operations are
not part of the data structure operation itself (as in predication),
but are executed by auxiliary (like wrappers in boosting).

As a simple example,
consider a non-negative counter, initially 0,
with \incr{} and \decr{} methods.
\incr{} does not return a value,
but \decr{} returns an error flag warning of an attempt to
decrement the counter below 0. 
We use just one STM memory location $\ell_0$,
and we associate the following read/write instructions with method operations:
\begin{itemize*}
  \item \var{incr()}: $read(\ell_0)$ whenever the counter is below 2.
  \item \var{decr()}: $write(\ell_0)$ whenever the counter is below 2.
\end{itemize*}
Consider a few cases:
(1) If the counter value is 52,
and there are concurrent \incr{} and \decr{} calls,
neither will access $\ell_0$ and the STM will detect no conflict.
The STM detects no conflict,
reflecting the absence of an abstract-level conflict.
(2) If the counter value is 0, 
and there are two concurrent \incr{} calls,
both will read $\ell_0$.
The STM detects no conflict,
reflecting the absence of an abstract-level conflict.
(3) If the counter value is 1,
and there are two concurrent \decr{} calls,
then both will write to $\ell_0$,
and the STM will detect a conflict.
The behavior is correct because the abstract operations do not
commute in this state: the first \decr{} will return normally,
and the second will report an error.

This example is nearly trivial, and requires only a single STM location.
There are many alternative conflict abstractions even for such a
simple structure.
The designer has wide latitude in deciding how many STM locations to
use, when to read, when to write, etc.
Also notice that the values written/read to/from STM locations doesn't actually matter
matter, as long as values written are unique,
such as sequence numbers or timestamps.

Here is a more complicated example.
Two hashmap operations commute if their keys are distinct:
\var{get}(42) commutes with \var{put}(11).
Therefore, a simple hashmap conflict abstraction is to assign
one STM location for every possible key,
and ensure that \var{get}($k_1$) causes a write to STM location $\ell_{k_1}$
and \var{put}($k_2$) causes a write to STM location $\ell_{k_2}$.
In this way if, say, a \var{get}(13) and \var{put}(13,\var{foo}) are
executed concurrently,
the STM will detect a conflict at location $\ell_{13}$.
Of course,
if there are many potential keys,
it is more sensible to allocate only $M$ locations,
for some $M$ of reasonable size,
and to have operations with key $k$ read and write to location $k \bmod M$.
This practice is similar to \emph{lock striping}~\cite{HerlihyS2008}.

Naturally,
there is a large design space for crafting such conflict abstractions,
and the proper choice depends on anticipated workload, architecture,
and other considerations..
In Section~\ref{section:eval} we explore a few conflict abstractions for
various data structures
and report on performance.

\paragraph{Correctness.} 
We now define what it means for a conflict abstraction to be correct.
We also briefly discuss how existing
software verification tools can verify the correctness of a conflict
abstraction against a given data structure specification/model.
Note that, to reason about correctness,
we \emph{do not need} the actual implementation of the thread-safe
concurrent objects.
Instead, it is sufficient to work with a \emph{model} (or sequential
implementation) of the abstract data type.

Conflict abstractions are represented as mathematical functions
that indicate which STM locations should be read or written for a
given data structure method $m(\bar{x})$
\[\left\{\begin{array}{rclcrcl}
f^{m,rd}_1 &:&\bar{x} \rightarrow \sigma \rightarrow {\cal B}
 &\;\;,\;\;\dots\;\;,\;\;&
   f^{m,rd}_M &:&\bar{x} \rightarrow \sigma \rightarrow {\cal B} \\
   f^{m,wr}_1 &:&\bar{x} \rightarrow \sigma \rightarrow {\cal B}
 &\;\;,\;\;\dots\;\;,\;\;&
   f^{m,wr}_M &:&\bar{x} \rightarrow \sigma \rightarrow {\cal B}\\
\end{array}\right\}\]
Above $f^{m,rd}_1$ is a function that consults the arguments $\bar{x}$ and
some information about the current state $\sigma$ of the data structure (e.g. whether
the counter value is below 2). The output of the function is a Boolean
indicating whether STM location $\ell_1$ should be read. Functions
$f^{m,wr}_1,f^{m,rd}_2,f^{m,wr}_2,$ etc are similar.

\begin{definition}[Conflict Abstraction]\label{def:ca}
A conflict abstraction
is correct provided that,
if  $m(\bar{\alpha})$ and $n(\bar{\beta})$ do not commute, then 
or any two method invocations $m(\bar{\alpha})$ and $n(\bar{\beta})$
(where $\bar{\alpha}$ and
$\bar{\beta}$ are the values of the parameters (actuals)),
their respective conflict abstractions will cause them to perform
conflicting STM memory access. That is, there will be some
memory location $\ell_i$ such that either:
\begin{enumerate*}
  \item $f^{m,rd}_i(\bar{\alpha},\sigma) = \textsf{true}$ and $f^{n,wr}_i(\bar{\beta},\sigma) = \textsf{true}$, or
  \item $f^{m,wr}_i(\bar{\alpha},\sigma) = \textsf{true}$ and $f^{n,rd}_i(\bar{\beta},\sigma) = \textsf{true}$, or
  \item $f^{m,wr}_i(\bar{\alpha},\sigma) = \textsf{true}$ and $f^{n,wr}_i(\bar{\beta},\sigma) = \textsf{true}$.
\end{enumerate*}
\end{definition}

The question of correctness can be reduced to satisfiability.
If the model is written in a format that can be understood by
SAT/SMT tools, then we can use these tools for verification.
For lack of space, this discussion continues in Appendix~\ref{apx:smt}.

%% file: shadow.tex
\section{Shadow Copies}
\label{section:shadow}
We now present techniques for lifting linearizable objects into a transactional setting using lazy updates.
A key challenge is that a transaction must be able to observe the results of speculative updates to shared objects, without those updates becoming visible to other transactions until the commit succeeds.

Previous approaches such as boosting~\cite{HerlihyK2008} relied on inverse operations to cleanup an aborted transaction, allowing updates to be made eagerly.
Unfortunately, this approach was coupled with pessimistic conflict resolution, where execution would wait when a conflict was detected.
In an optimistic setting, transactions execute as if they will not encounter conflicts, and abort or retry when they are eventually detected.
If these are detected early enough, the eager update strategy still works, but this approach is not compatible with all STMs, and so lazy updates may be preferable in many situations.

\paragraph{Replay Wrappers}
To implement lazy updates, we use \emph{replay wrappers} to queue updates, and only apply them when it is known the transaction will commit.
The wrappers channel pending ADT operations (\texttt{put}(5,''foo''), \texttt{get}(23), etc.) into a log associated with each transaction.
When it is known that a transaction will commit, its log entries are applied atomically, behind the STM's native locking mechanisms.
If, instead, the transaction aborts, its log is dropped.

\paragraph{Return Values}
When an operation is logged, the transaction which executed it must also be able to obtain the value returned by that operation (e.g. what is the result of \texttt{set(23,5)}?).
To obtain these values, transactions must utilize a \emph{shadow copy} of that data structure.
We define two different approaches for implementing such functionality.

\paragraph{Memoization.}
For some data-structures (e.g. sets or maps), the results of an operation (even an update) can be computed purely from the initial state of the wrapped data-structure, or from the arguments to other pending operations.
In these cases, we may implement shadow copies by memoization.
Repeated operations to the same key can be cached in a transaction-local table, and queried, to determine the results of the next operation on that key.
If they key is not present, it's stated can be determined by reading the unmodified backing data structure.

We implemented this approach in our \textsf{LazyHashMap}, using Java's
\textsf{ConcurrentHashMap} as the underlying data-structure.

\paragraph{Snapshots.}
For many data structures, memoization will be insufficient.
A more general approach uses the fast-snapshot semantics provided by many concurrent data structures.~\cite{petrank2013lock,ProkopecSnapQueue,BronsonPCBST,ProkopecCTrie}
The first time a transaction attempts to perform an update, a snapshot is made, and all further updates are performed on that snapshot.
Whenever a transaction commits, any changes to the snapshot are replayed onto the shared copy.

We implemented two data-structures this way: \textsf{LazyTrieMap} and \textsf{LazyPriorityQueue}.
For the LazyTrieMap, we used Scala's concurrent \textsf{TrieMap} as the underlying data-structure.
For the LazyPriorityQueue, we designed a new base copy-on-write data structure.

%% file: opacity.tex
\section{Opacity}
\label{section:opacity}
\emph{Opacity}~\cite{opacity} is a correctness condition that 
(simplifying somewhat)
ensures that committed transactions appear to have executed in a serial order,
and that aborted transactions observed consistent memory states at all times.
Modern STMs typically guarantee opacity.

In this section we provide proof sketches of the intuitive connection from
the \emph{Push/Pull} formal semantic model~\cite{KP:PLDI2015}
to show that transactional objects
constructed using the Proust framework satisfy opacity.
The challenge is how to integrate conflict abstractions
(Section~\ref{section:conflict}),
shadow copy objects (Section~\ref{section:shadow}),
eager-vs-lazy/optimistic-vs-pessimistic variants, and
differing assumptions about the guarantees of the underlying STM.
Below we state the theorems. For lack of space, proofs can be found in
Appendix~\ref{apx:proofs}.

A transaction $T$ consists of a sequence of data structure operations $m_1,m_2,...,m_n$.
For simplicity, we model state as a single array \var{mem} that has
two \emph{disjoint} regions:
$\var{mem}[\alpha,\beta,...]$ for conflict abstractions and
$\var{mem}[x,y,...]$ is for shared variable references.
Both are managed by the same underlying STM.
Finally, for each operation $m_i$, the conflict abstraction
associated with $m_i$ is denoted $CA(m_i) \equiv \alpha_i^1,\alpha_i^2,\dots,\alpha_i^k$.

\begin{theorem}\label{thm:pess} Pessimistic Proust is opaque.
  \begin{proof}(Sketch) See Appendix~\ref{apx:proofs}.
   Based on~\cite{HerlihyK2008,KP:PLDI2015}. 
Transactions acquire locks enforcing commutativity of operations on the Proustian objects, and releases them only after the transaction commits.
Updates, whether performed eagerly or lazily, can not be observed until these locks are released. Thus opacity is satisfied.
  \end{proof}
\end{theorem}

\begin{theorem}\label{thm:opteager} Eager/Optimistic Proust is opaque if the STM detects all conflicts eagerly.
  \begin{proof}(Sketch) See Appendix~\ref{apx:proofs}.
For non-commutative operations $m_i$ and $n_j$ from
different uncommitted transactions, there will be some memory location
$\gamma \in CA(m_i) \cap CA(n_j)$ (Definition~\ref{def:ca}).
WLOG, $m$ performs a \var{write} to $\alpha^1_i,\alpha^2_i,\dots\in CA(m_i)$
before performing operation $m_i$, and $n$ either \var{read}s or \var{write}s  to $\alpha^1_j,\alpha^2_j,\dots\in CA(n_j)$
Therefore, $m$ and $n$ will perform conflicting operations on \var{mem}[$\gamma$], and at least one will aborted.
Aborted transaction perform their recorded inverse operations and---following the same argument used for boosting~\cite{HerlihyK2008,KP:PLDI2015}---the
commutative nature of the operations performed thus far ensures that
the inverses will not be visible to concurrent transactions.
  \end{proof}
\end{theorem}
\begin{theorem}\label{thm:optlazy} Lazy/Optimistic Proust is opaque.
  \begin{proof}(Sketch) See Appendix~\ref{apx:proofs}.
Shadow copies permit active transactions to ``clone'' their
own replica of the current committed version, but it is not a full
``clone'' because the shadow copy may rely on the latest
committed version to generate return values.
To mitigate this, we surround execution of a Proustian operation $m_i$
with the relevant conflict abstraction operations $CA(m_i)$ as:
\[\begin{array}{l}
\textsf{foreach } \alpha \in CA(m_i) \textsf{ do write}(\alpha) \textsf{ done;}
\;\;\;\;\;\;
m_i();
\;\;\;\;\;\;
\textsf{foreach } \alpha \in CA(m_i) \textsf{ do read}(\alpha) \textsf{ done;}\\
\end{array}\]
Together, these sets of \var{write}s and \var{read}s check whether
some other transaction $T'$ has committed a conflicting operation $n_j$.
During the commit of $T'$, the underlying STM will grant $T'$ exclusive
access to all \var{mem} locations accessed by $T'$, including the
regions of \var{mem} for references and for conflict abstraction locations.
The \var{write} operations above announce $m_i$ via the conflict
abstraction operations on \var{mem}. The \var{read} operations, if
they do not trigger an abort,
ensure that the transaction is working with a shadow copy
that has not been recently invalidated due to $T'$ committing a conflicting $n_j$
(whose conflict abstraction $CA(n_j)$ overlaps with $CA(m_i)$).
  \end{proof}
\end{theorem}

The consequence of these careful algorithms is that, despite lazy conflict
detection, it is still impossible for a transaction to observe the
effects of another concurrent uncommitted transaction.

%% file: implementation.tex
\section{\toolName{} Implementation Details}
\label{sec:impl}
In addition to developing the methodology and techniques described so far, we developed a library implementation of Proust on top of ScalaSTM, which we call \toolName.

\toolName{} provides an API implementing abstract locks, replay logs for both shadow-copy techniques, and lock allocator policies of both the optimistic\footnote{We note that the default backend for ScalaSTM is CCSTM which uses a mixed conflict detection strategy. ~\cite{bronson2010ccstm} One consequence of this is that eager/optimistic objects will not satisfy opacity out of the box.
We nonetheless felt it was useful to provide them as ScalaSTM supports pluggable backends, and some practical applications may endure opacity violations without ill-effect.} and pessimistic varieties.
It also provides a number of wrapped Proustian data structures out of the box, which can be used as-is, or serve as example code for developers to create their own wrappers.

In this section we will outline the core features of \toolName, and highlight ways in which the expressive power it offers is superior to that of earlier works.

As a first example we consider the implementation of a transactional Map API (Listing \ref{lst:txnMapTrait}) against the AbstractLock interface of Listing \ref{lst:abstractLockAPI}.
Figures \ref{fig:eagerMap} and \ref{fig:lazyMap}  depict the eager and lazy implementations, respectively.
They are virtually identical, except that in the latter, no inverses are needed, and instead each operation on \texttt{map} is forwarded through a call to \texttt{log}.
We note that the choice of optimistic or pessimistic conflict resolution remains with the LockAllocatorPolicy used in the constructor.
\begin{figure*}
\begin{subfigure}[b]{3.15in}\centering
\begin{lstlisting}[language=Scala,basicstyle=\scriptsize]
class TrieMap[K,V](val lAP:LockAllocatorPolicy[K]) extends MapTrait[K, V] {
	private val map = RawTrieMap[K,V]()
	val uStrat = Eager
	           
	def put(key: K, value: V)(implicit txn:InTxn) = abstractLock(Write(key)){
		val ret = map.put(key, value)
		if(ret.isEmpty) committedSize() += 1
		ret
	}{_.map(map.put(key, _)).getOrElse(map.remove(key))}
	def get(key: K)(implicit txn:InTxn) = abstractLock(key){map.get(key)}()	
	def contains(key: K)(implicit txn:InTxn) = abstractLock(key){map.contains(key)}()
	def remove(key: K)(implicit txn:InTxn) = abstractLock(Write(key)) {
		val ret = map.remove(key)
		if(ret.isDefined) committedSize() -= 1
		ret
	}{_.foreach{map.put(key, _)}}
}
\end{lstlisting}
\caption{\label{fig:eagerMap} An eager implementation of a Proustian
  Map.}
\end{subfigure}
\quad
\begin{subfigure}[b]{3.15in}\centering
\begin{lstlisting}[language=Scala,basicstyle=\scriptsize]
class LazyTrieMap[K,V](val lAP:LockAllocatorPolicy[K]) extends MapTrait[K, V] {
	...
	val uStrat = Lazy
	private val log = ReplayLog.construct[RawTrieMap[K,V]](new SnapshotReplay(_, _.snapshot), map)
	private def readOnly[Z](f:RawTrieMap[K,V] => Z)(implicit txn:InTxn) = {
		ReplayLog.readOnly(replays, f, map)
	}                   
	def put(key: K, value: V)(implicit txn:InTxn) = abstractLock(Write(key)){
		val ret = (log()()= _.put(key, value)))
		if(ret.isEmpty) committedSize() += 1
		ret
	}()
	def get(key: K)(implicit txn:InTxn) = abstractLock(key){readOnly(_.get(key))}()
	...
}
\end{lstlisting}
\caption{\label{fig:lazyMap} A lazy implementation of a Proustian
  Map.}
\end{subfigure}
  \caption{ Eager and lazy implementations of a Proustian Map.  In the
    lazy implementation, \texttt{ReplayLog.construct} returns a
    \texttt{TxnLocal} that allocates a new log the first time the Map
    is written during each transaction. \texttt{readOnly} provides an
    optimization to avoid initializing the log until it is known that
    a replay is actually necessary. (Imports have been elided.)}
\end{figure*}

As a more nuanced example, we next consider the problem of wrapping a Priority Queue, using the interface in Listing \ref{lst:txnPQueueTrait}, which provides categorizes all operations by their effect on either \texttt{PQueueMin} or \texttt{PQueueMultiSet}.
The authors of Transactional Boosting note that for a priority queue all calls to \texttt{add(x)} commute, and that \texttt{add(x)} commutes with \texttt{removeMin()/y} if $y \le x$; however, in their implementation they are unable to expressly this commutative relationship cleanly, and instead approximate it conservatively with a read/write lock.~\cite{HerlihyK2008}
The situation gets messier when you consider additional operations, they discuss \texttt{min} incompletely, and provide no implementation.
Other useful common operations (e.g. \texttt{contains} and \texttt{size}) are out of the picture entirely.

\begin{wrapfigure}[15]{r}{3.1in}
\centering
\begin{lstlisting}[language=Scala,basicstyle=\scriptsize]
	def insert(v : Value)(implicit txn:InTxn) : Unit = {
		abstractLock(Write(PQueueMultiSet), 
				min.collect{ 
					case curM if v < curM => Write(PQueueMin)
				}.getOrElse{Read(PQueueMin)}){
			val wrapper:LazyDeletion = v
			pQueue.add(wrapper)
			committedSize() += 1
			wrapper
		}{_.delete}
	}
\end{lstlisting}
\caption{\label{fig:txnPQueue} An insert implementation for PriorityQueue (using the Java standard library's \texttt{PriorityBlockingQueue} as a backing data structure).}
\end{wrapfigure}
By characterizing operations explicitly in terms of abstract state, we are able to express commutativity with a number of rules which is linear in the state space, rather than quadratic in the number of methods.
These rules are as follows: PQueueMin allows multiple readers and a single writer, whereas PQueueMultiSet allows multiple writers or multiple readers (but not both simultaneously).
Since \texttt{AbstractLock}'s first argument list is evaluated like any other Scala expression we can express our \texttt{insert} as shown in Figure \ref{fig:txnPQueue} (using the same lazy-deletion trick utilized in the Boosting paper).

This example also highlights another shortcoming in the original Boosting methodology: eager updates don't mix well with data-structures whose operations don't have efficient inverses.
Proustian methodology on the other hand allows us to utilize a lazy update strategy instead\footnote{We are not aware of any publicly available implementations of concurrent heaps or ordered maps which support efficient snapshot operations; however, \toolName{} contains an experimental implementation that uses copy-on-write semantics, and the broader area of snapshot-able collections is an area of active research.~\cite{petrank2013lock,ProkopecSnapQueue,BronsonPCBST,ProkopecCTrie}}.

%% file: experiment.tex
\section{Evaluation}
\label{section:exp}
\label{section:eval}

We evaluated several of our map implementations for \toolName{} with a benchmarking setup similar to that used by Bronson, et al. for predication~\cite{BronsonCCOn2010}.
For each experiment, we performed $1000000$ randomly selected operations on a shared map, split across $t$ threads, with $o$ operations per transaction.
A $u$ fraction of the operations were writes (evenly split between \texttt{put} and \texttt{remove}), and the remaining $(1- u)$ were \texttt{get}.
We varied $1 \le t \le 32$, $1 \le o \le 256$, and $u \in \left\{0,0.25,0.5,0.75,1\right\}$.
For each configuration, we warmed up the JVM for $10$ executions, then timed each of the following $10$ executions, garbage collecting in between to reduce jitter, and reported the mean and standard deviation.

Unlike the predication paper, we did not vary the key range from which operations were randomly selected, using instead a fixed value of $1024$.
The reason for this difference in methodology was that techniques for garbage collecting the synchronization primitives used by abstract locks was not a focus of this paper; however,
the techniques developed by Bronson, et al. for garbage collecting predicates should be directly applicable.
\begin{figure*}
\centering
\small
\setlength\tabcolsep{2.5pt}
\begin{tabular}{cl|cccc}
& & \multicolumn{4}{c}{Fraction of operations that are writes ($u$)}\\
& & 0.25 & 0.5 & 0.75 & 1 \\
\hline
\multirow{9}{*}{\rotatebox{90}{Operations per transaction ($o$) \hspace{18em}}} & & \multicolumn{4}{c}{\textsf{MapThroughputTest}} \\
&
\rotatebox{90}{\hspace{.5in} 1} &
\includegraphics[width=1.45in,trim={0.75cm 0.25cm 0.75cm 0},clip]{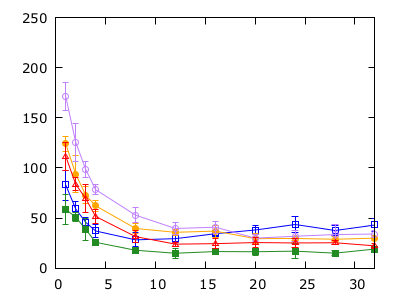} &
\includegraphics[width=1.45in,trim={0.75cm 0.25cm 0.75cm 0},clip]{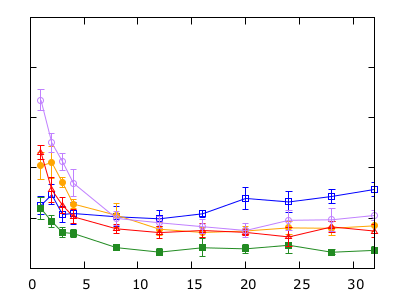} &
\includegraphics[width=1.45in,trim={0.75cm 0.25cm 0.75cm 0},clip]{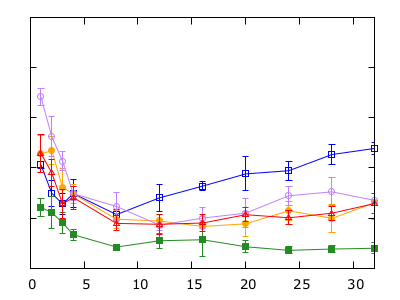} &
\includegraphics[width=1.45in,trim={0.75cm 0.25cm 0.75cm 0},clip]{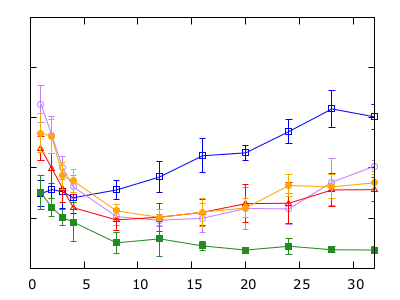} \\
&
\rotatebox{90}{\hspace{.5in}2} &
\includegraphics[width=1.45in,trim={0.75cm 0.25cm 0.75cm 0},clip]{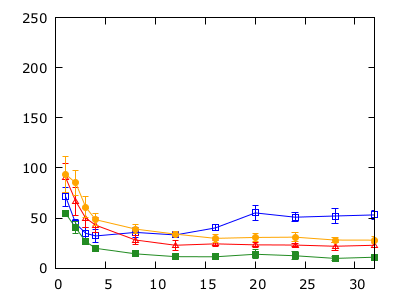} &
\includegraphics[width=1.45in,trim={0.75cm 0.25cm 0.75cm 0},clip]{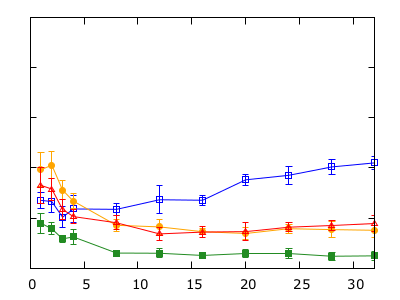} &
\includegraphics[width=1.45in,trim={0.75cm 0.25cm 0.75cm 0},clip]{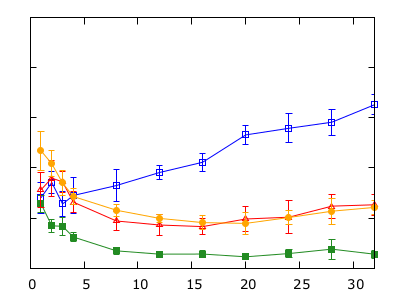} &
\includegraphics[width=1.45in,trim={0.75cm 0.25cm 0.75cm 0},clip]{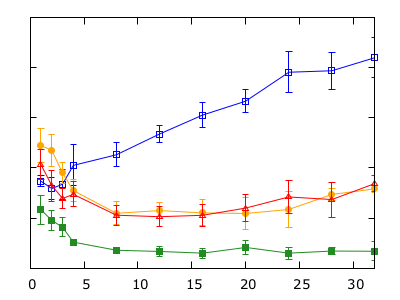} \\
&
\rotatebox{90}{\hspace{.5in}16} &
\includegraphics[width=1.45in,trim={0.75cm 0.25cm 0.75cm 0},clip]{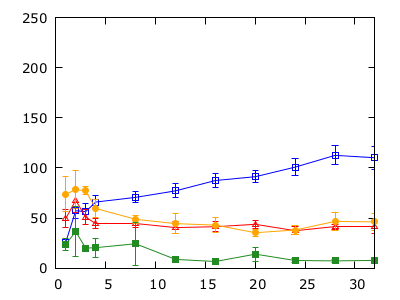} &
\includegraphics[width=1.45in,trim={0.75cm 0.25cm 0.75cm 0},clip]{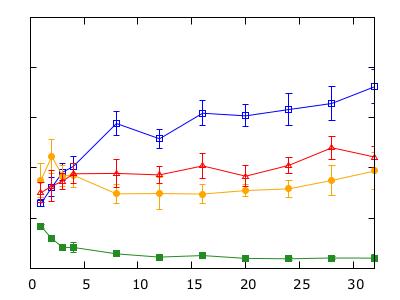} &
\includegraphics[width=1.45in,trim={0.75cm 0.25cm 0.75cm 0},clip]{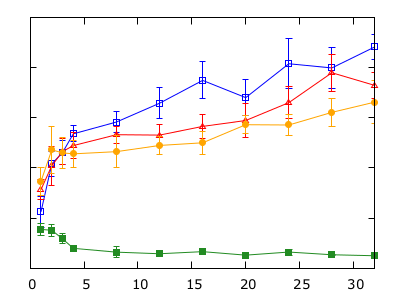} &
\includegraphics[width=1.45in,trim={0.75cm 0.25cm 0.75cm 0},clip]{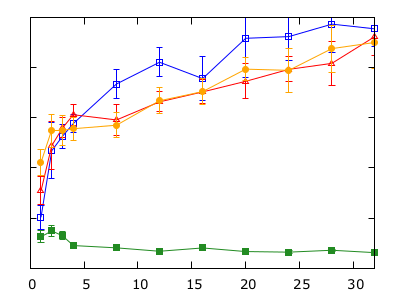} \\
&
\rotatebox{90}{\hspace{.5in}256} &
\includegraphics[width=1.45in,trim={0.75cm 0.25cm 0.75cm 0},clip]{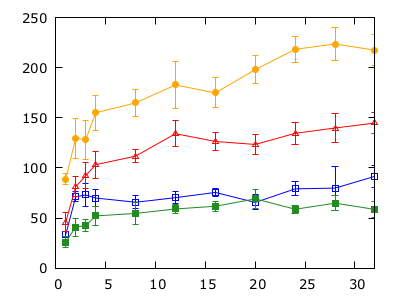} &
\includegraphics[width=1.45in,trim={0.75cm 0.25cm 0.75cm 0},clip]{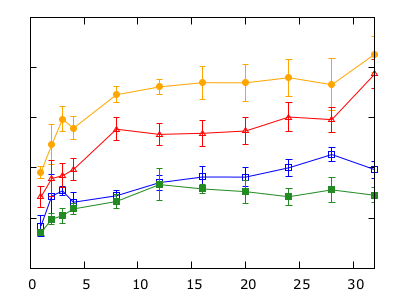} &
\includegraphics[width=1.45in,trim={0.75cm 0.25cm 0.75cm 0},clip]{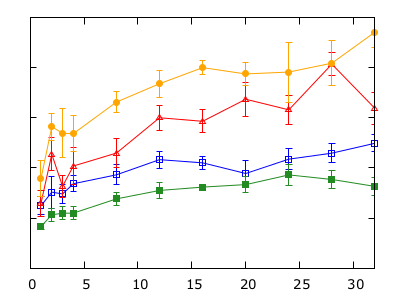} &
\includegraphics[width=1.45in,trim={0.75cm 0.25cm 0.75cm 0},clip]{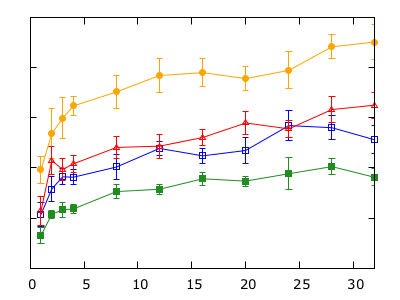} \\
& & \multicolumn{4}{c}{\includegraphics[width=4.0in]{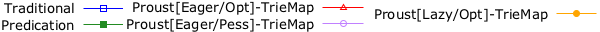}}\\
\cline{3-6}
& & \multicolumn{4}{c}{\textsf{MapThroughputTest} - Memoizing Shadow Copies} \\
&
\rotatebox{90}{\hspace{.5in}256} &
\includegraphics[width=1.45in,trim={0.75cm 0.25cm 0.75cm 0},clip]{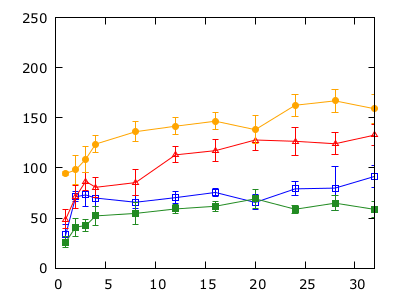} &
\includegraphics[width=1.45in,trim={0.75cm 0.25cm 0.75cm 0}]{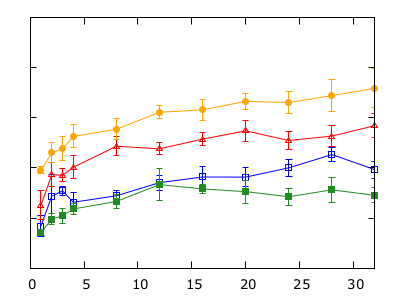} &
\includegraphics[width=1.45in,trim={0.75cm 0.25cm 0.75cm 0}]{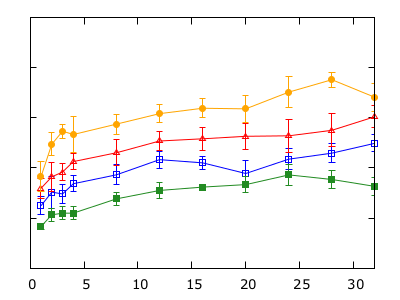} &
\includegraphics[width=1.45in,trim={0.75cm 0.25cm 0.75cm 0}]{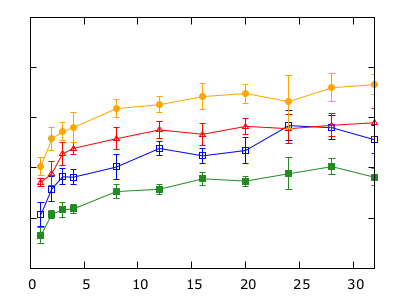} \\
& & \multicolumn{4}{c}{\includegraphics[width=3.5in]{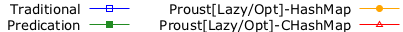}}\\
\end{tabular}
\caption{\label{fig:mapThroughput} Time to process $1000000$ operations on concurrent maps using a 32-core Amazon EC2 \texttt{m4.10xlarge} instance as the number of threads increases.  Each chart is the result for a particular fraction of writes and operations per transaction.  For each chart, the x-axis is the number of threads from 0 to 32 and the y-axis is the average time in milliseconds from 0 to 250.}
\end{figure*}

The experimental results depicted in Figure \ref{fig:mapThroughput}
display the effects of several competing trends.  We ran our
experiments on an Amazon EC2 \texttt{m4.10xlarge}
instance\footnote{\url{https://aws.amazon.com/blogs/aws/the-new-m4-instance-type-bonus-price-reduction-on-m3-c4/}},
which has 40 vCPUs and 160 GB of RAM.

Two notes about the experimental setup. 
First, we were only able to achieve a weak coupling with the CCSTM contention manager for the pessimistic experiments, and found that under the artificially high contention seen in these experiments, longer transaction times could lead to live-lock, as the STM successfully resolved cyclic locking dependencies, but was unable to instruct the transaction attempts to back off without access to information about the instigating (non-STM) memory accesses.
For this reason we only show the pessimistic results in the initial $o=1$ experiments.
Second, even though the Eager/Optimistic configuration does not satisfy opacity under the CCSTM backend for ScalaSTM, we chose to benchmark it anyway, and did not observe any instances where this violated correctness (notably our benchmark makes no explicit control flow decisions based on the results of map accesses, and ScalaSTM performs an abort and retry if it ever observes an unchecked exception).
It does seem likely that some performance penalty was paid for late detection of inconsistent memory accesses, and we believe this speaks well to the potential performance of Eager/Optimistic Proustian data structures on STMs where they satisfy opacity.

Intuitively, the performance of the Proustian maps scales much better than the traditional STM implementation as contention increases, due to varying $t$ and $u$ (though we are consistently outperformed by the highly engineered predication implementation); however, increasing values of $o$ have a negative influence on the relative performance of the Proustian maps.
Intuitively, this is to be expected as applying our logs (either to undo in the eager setting, or to replay in the lazy setting) takes time proportional to the number of updates performed, whereas predication and traditional implementations replay with time proportional to the number of unique memory locations updated, and as $o$ increases, so does the probability that multiple writes will alter the same location.

An optimization available to Proustian data structures that utilize a memoizing shadow copy rather than a functional snapshot is to replay synthetic updates to apply only the final state of each abstract state element. The effects of this optimization are seen at the bottom of Figure~\ref{fig:mapThroughput}.

%% file: relwork.tex
\section{Related Work}
Early work on exploiting commutativity for concurrency control
includes Korth~\cite{Korth1983}
Weihl~\cite{weihlcommu},
CRDTs~\cite{Shapiro2011},
and Galois~\cite{Kulkarni2009}.
Some false conflicts in STMs can be alleviated by escape mechanisms such as
open nesting~\cite{opennested}
elastic transactions~\cite{felber2009},
transactional collection classes~\cite{Carlstrom2007},
but these mechanisms can be complex and difficult to use correctly.
Other mechanisms that exploit commutitivity for STM systems include
boosting~\cite{HerlihyK2008}, automatic semantic locking~\cite{Golan-Gueta2014},
STOs~\cite{HermanIHTKLS2016}, and predication~\cite{BronsonCCOn2010}.
Dimitrov \emph{et al.}~\cite{DimitrovRVK2014} described a method for translating
commutativity formulae into so-called \emph{access point specifications} that
could be used by a dynamic race detection tool.

%% file: proofs.tex
\section{Opacity Proof Sketches}
\label{apx:proofs}

Here we sketch the proofs of Theorems~\ref{thm:pess}, \ref{thm:opteager}, and~\ref{thm:optlazy}.
We use the \emph{Push/Pull} formal semantic model~\cite{KP:PLDI2015}.

A transaction $T$ consists of a sequence of data structure operations $m_1,m_2,...,m_n$.
For simplicity, we model state as a single array \var{mem} that has
two \emph{disjoint} regions:
$\var{mem}[\alpha,\beta,...]$ for conflict abstractions and
$\var{mem}[x,y,...]$ is for shared variable references.
Both are managed by the same underlying STM.
Finally, for each operation $m_i$, the conflict abstraction
associated with $m_i$ is denoted $CA(m_i) \equiv \alpha_i^1,\alpha_i^2,\dots,\alpha_i^k$.

\paragraph{Background: Push/Pull.}

The Push/Pull model provides a general framework for aligning
transactional memory algorithms with
serializability/opacity arguments (via simulation relations)~\cite{KP:PLDI2015}.
The model has no concrete state but instead
consists of a single shared global log of operations,
as well as per-transactional local logs.
There are a few simple rules---named \PUSH{}, \PULL{}, etc.---that ferry (the names of)
ADT operations between these logs in ways that correspond to natural stages in a
transactional memory algorithm.
Different algorithms (pessimistic, optimistic, dependent transactions,
etc.)  combine these rules in different ways.
Here are some example stages.
A transaction perhaps begins by \PULL{}ing operations from the shared log
into its local log,
viewing effects that other transactions previously shared.
It then may take a step forward, appending a new operation $m$ to its local log via the \APPLY{}
rule.
Next, it may \PUSH{} this $m$ to the shared log, publishing this effect.
At this stage, the transaction may
not have committed (\eg\ when modeling boosting~\cite{HerlihyK2008},
commutative operations are \PUSH{}ed).
Meanwhile, the \PULL{} rule enables transactions to update
their local view with operations that are permanent (that is, they
correspond to committed transactions) or even to view the effects of
another uncommitted transaction (\eg, for eager conflict
detection~\cite{dstm} or dependent transactions~\cite{ut}).
One can also move in the reverse direction:
Push/Pull includes an \UNPULL{} rule which discards a
transaction's knowledge of an effect due to another thread, and an
\UNPUSH{} rule which removes a thread's operation from the shared
view, perhaps implemented as an inverse. The \UNAPP{} rule is useful
for rewinding a transaction's local state.  Finally, there is a simple
commit rule \CMT{} that, roughly, stipulates that all operations must
have been \PUSH{}ed and all \PULL{}ed operations must have been
committed.
(For lack of space, in the descriptions below, we will omit some detail
about how aborting is modeled with \UNPULL{} and \UNAPP{}.)

\paragraph{Proof of Theorem~\ref{thm:pess}: Pessimistic Proust.}
In the pessimistic variant, conflict detection on Proustian
objects is done eagerly using abstract locks (as
in~\cite{HerlihyK2008}). In the Push/Pull model, this means that
transactions \APPLY{} locally, but do not \PUSH{} these effects
until they are sure the effects commute with all other
uncommitted operations that were \PUSH{}ed by other threads.
When another transaction commits, it is guaranteed not to conflict due to
the restriction that all \PUSH{}ed operations must commute.
It was shown~\cite{KP:PLDI2015} that this algorithm satisfies opacity.

Memory references $\var{mem}[x,y,...]$ may be
treated eagerly or lazily, depending on the underlying STM.
Ultimately, in either case, serializability is ensured because
all memory and Proustian object operations must be \PUSH{}ed
(which involves a conflict check) before \CMT{}.
With an eager STM, we model the STM's early knowledge of conflict
between memory references by \PUSH{}ing them to the shared log
immediately after they have been \APPLY{}ed locally.
Lazy memory references are modeled as transactions that
\PUSH{} Proustian object operations, but not \PUSH{}ing
memory operations until just before a commit. In this case, transactions
still cannot observe the effects of uncommitted transactions because
uncommitted \PUSH{}ed operations commute, and
reference variable have local-only views.

\newcommand\anno[1]{$\langle\!\!\langle$#1$\rangle\!\!\rangle$}
\paragraph{Proof of Theorem~\ref{thm:opteager}: Eager Optimistic Proust.}
Here we use \emph{conflict abstractions} (Section~\ref{section:conflict})
so that Proustian object conflict is detected by the underlying STM.
Recall that a conflict detected in \var{mem}[$\alpha,\beta,...$]
means that two Proustian object operations do not commute.
In what follows, we annotate the elements of the algorithm with
their corresponding Push/Pull rule using the notation \anno{...}.

As shown in Figure~\ref{fig:designspaces}, the key requirement for
this version is that the
underlying STM detects (read/write \emph{and} write/write) conflicts
eagerly. Given the definition of conflict abstraction  (Definition~\ref{def:ca}),
if there are non-commutative operations $m_i$ and $n_j$ from
different uncommitted transactions, there will be some memory location
$\gamma \in CA(m_i) \cap CA(n_j)$.
In our eager implementation a transaction first performs a \var{write}
operation on memory locations $\alpha^1_i,\alpha^2_i,\dots\in CA(m_i)$
before performing operation $m_i$
\anno{ \APPLY{}($m_i$) and try to \PUSH{}($m_i$) }.
Therefore, both transactions will attempt
to write \var{mem}[$\gamma$]. Due to the eager nature,
this \var{mem}[$\gamma$] conflict will be detected and the second of the two 
transactions will be aborted before it performs its Proustian operation
\anno{ failing to \PUSH{}($m_i$) }.
The aborted transaction will perform inverse operations
\anno{ invoking \PUSH{}($m_{i-1}$);\UNAPP{}($m_{i-1}$), etc in reverse order}
and---following the same argument used for boosting~\cite{HerlihyK2008,KP:PLDI2015}---the
commutative nature of the operations performed thus far ensures that
the inverses will not be visible to concurrent transactions.
If a transaction commits,
the underlying STM will ensure that
writes to both regions of \var{mem} will become permanent,
with no effect on concurrent transactions whose reads/writes to \var{mem} are
treated with eager conflict detection \anno{ \CMT{} }.
Since this variant operates directly on the Proustian objects,
the committed effects are immediately visible \anno{ other transactions
  \PULL{} all before each subsequent \APPLY{} }.

\paragraph{Proof of Theorem~\ref{thm:optlazy}: Lazy/Optimistic Proust.}
The third variant is designed to exploit an STM that performs \emph{lazy}
read/write and lazy write/write conflict detection (Figure~\ref{fig:designspaces}).
In these circumstances, since conflict is detected late in the game
(at commit time), it would
be dangerous to allow uncommitted transactions to perform mutation
operations directly on the shared state.

In response, shadow copy data structures
(Section~\ref{section:shadow}) allow transactions to proceed with
operations on Proustian objects (and observe their corresponding return
values) without exposing these effects to concurrent uncommitted
transactions.
Informally, shadow copies permit active transactions to ``clone'' their
own replica of the current committed version (using a couple of
different implementation strategies) \anno{\PULL{}($n$) for only committed $n$}.
It is not a full ``clone'' because the shadow copy may rely on the latest
committed version to generate return values.
Modeling this behavior, our lazy optimistic transactions
\PULL{} to retrieve the committed effects on a Proustian object
perform \APPLY{} rules to step forward but,
unlike the previous variants, do not perform \PUSH{} rules until it
has been determined (lazily) that the transaction can commit.

While transaction $T$ is executing, what happens if another transaction
$T'$ commits conflicting operations? Trouble arises
because  $T'$ replays its log on each Proustian object, updating
the object to a new committed version, possibly invalidating future $T$ operations: the
shadow copy belonging to $T$ may need to consult the underlying data structure
for return values and view the effects of $T'$.
In response, we surround execution of a Proustian operation $m_i$
with the relevant conflict abstraction operations $CA(m_i)$ as:
\[\begin{array}{l}
\textsf{foreach } \alpha \in CA(m_i) \textsf{ do write}(\alpha) \textsf{ done;}
\;\;\;\;\;\;
m_i();
\;\;\;\;\;\;
\textsf{foreach } \alpha \in CA(m_i) \textsf{ do read}(\alpha) \textsf{ done;}\\
\end{array}\]
Together, these sets of \var{write}s and \var{read}s check whether
some other transaction $T'$ has committed a conflicting operation $n_j$.
During the commit of $T'$, the underlying STM will grant $T'$ exclusive
access to all \var{mem} locations accessed by $T'$, including the
regions of \var{mem} for references and for conflict abstraction locations
\anno{ $T'$ performs \PUSH{}($n_j$) for each operation $n_j$,
  \PUSH{}(read($x$))/\PUSH{}(write($x$)) for every accessed $x$,
  and then \CMT{} }.
The \var{write} operations above announce $m_i$ via the conflict
abstraction operations on \var{mem}. The \var{read} operations, if
they do not trigger an abort,
ensure that the transaction is working with a shadow copy
that has not been recently invalidated due to $T'$ committing a conflicting $n_j$
(whose conflict abstraction $CA(n_j)$ overlaps with $CA(m_i)$)
\anno{ Otherwise, $T$ does not perform \APPLY{}($m_i$) and instead
invokes \UNAPP{}($m_{i-1}$), etc in reverse order}.

%% file: smt.tex
\section{Verification of Conflict Abstractions}
\label{apx:smt}

The question of correctness can be reduced to satisfiability.
If the model is written in a format that can be understood by
SAT/SMT tools, then we can use these tools for verification.
Here is an example SAT/SMT model of a counter:
\begin{center}
\begin{program}[style=tt]
(de\tab fine-fun incr ((c0 Int) (c1 Int)) Bool
    (= (+ c0 1) c1)) \untab
(de\tab fine-fun decr ((c0 Int) (c1 Int) (err Bool)) Bool
  (and \;\;\; (= (- c0 1) c1) \;\;\; (ite (< c1 0) (= err true) (= err false))))
\end{program}
\end{center}
The two functions above specify how \var{incr} and \var{decr}
respectively transform a counter from its current value \var{c0} into
its new value \var{c1} and, in the case of \var{decr}, whether
the \var{err} flag is raised.
One can also model more complicated data-structures such
as hashtables~\cite{bansal2015commutativity}.

To the
right is an example for \var{incr()}/\var{decr()}. Intuitively, 
the SMT begins by asserting the
\begin{wrapfigure}[9]{r}{3.0in}
{\footnotesize
\begin{program}[style=tt]
  (assert (incr\_CA l0 l1 c0))    ; incr tickles the STM
  (assert (incr c0 c1))              ; incr executes
  (assert (decr\_CA l1 l2 c1))    ; decr tickles the STM
  (assert (not (conflict l2)))       ; no conflict detected
  (assert (decr c1 c2 err1))         ; decr executes
  ; check that the other order:
  (assert (decr c0 c3 err2))
  (assert (incr c3 c4))
  (assert (or (not (= c2 c4)) (not (= err1 err2))))
  (check-sat)
\end{program}
}
\end{wrapfigure}
following constraints:
\begin{enumerate*}
  \item Method $m$ performs its conflict abstraction reads/writes.
  \item Method $m$ performs its data-structure operation (e.g. \var{incr()}).
  \item Method $n$ performs its conflict abstraction reads/writes.
  \item No read/write or write/write conflict occurs.
  \item Method $n$ performs its data-structure operation (e.g. \var{decr()}).
\end{enumerate*}
We now need to ensure that the resulting state is the same as it would
have been if the operations executed in the opposite order. Using different
variable names (e.g. \var{c3}, \var{err2}, etc.), we then assert the 
other order (\var{decr()}\ before \var{incr()}).
Finally, we assert that the results (return values and final state) were
different and check whether this is satisfiable.
If it is not satisfiable, then the conflict abstraction is correct.

\begin{theorem}
  If the above encoding is unsatisfiable, then the
  conflict abstraction is sound.
\end{theorem}

As is often the case, this automatic verification technique can be
used as a building-block for an automatic synthesis technique.
We believe that it would be interesting to apply the approach of counter-example guided inductive synthesis
(CEGIS)~\cite{GJTV:PLDI2011,IGIS:OOPSLA2010,BCKGM:PPoPP2013,solar2008},
using SAT/SMT counter-examples as the basis for constructing
$f^{m,rd}_1,f^{m,rd}_1,$ etc. We leave this to future work.